\documentclass[a4paper,12pt]{article}
\usepackage{epsfig}
\usepackage{amsmath}
\usepackage{amssymb}

\def\today{\number\day
	   \space\ifcase\month\or
	     January\or February\or March\or April\or May\or June\or
	     July\or August\or September\or October\or November\or December\fi
	   \space\number\year}
\def\beq{\begin{equation}}
\def\eeq{\end{equation}}
\def\bea{\begin{eqnarray}}
\def\eea{\end{eqnarray}}

\def\ifb{${\rm fb}^{-1}$}

\def\ra{\rightarrow}

\def\lra{\leftrightarrow}
\def\tw{$t \, W^-~$}
\def\ttbar{$t \, \bar{t}~$}
\def\wwj{$W^+ \, W^- \, j~$}
\def\wwb{$W^+ \, W^- \, b~$}
\def\wwc{$W^+ \, W^- \, c~$}

\def\alphas{\alpha_{S}}

\begin{document}

\thispagestyle{empty}
\hspace*{-0.65cm}\today \hfill  ANL-HEP-PR-99-98 \\
hep-ph/9909352

\vspace{1.5cm} 

\begin{center}
{ {\Large {\bf The \tw Mode of Single Top Production}}}

\vspace{1.5cm} 

\centerline{ {\rm Tim~~M.~P.~~Tait\footnote{Electronic address:
{tait}@anl.gov}} }
\vspace{1cm}
{\it Argonne National Laboratory \\
Argonne, Illinois 60439, USA} \\

\vspace{1.5cm}

\begin{abstract}
\noindent 
The \tw mode of single top production is proposed as an 
important means to study the weak interactions of the 
top quark.  While the rate of
this mode is most likely too small to be observed at 
Run~II of the Fermilab Tevatron, it is expected to be 
considerably larger at the CERN LHC.  In this article the inclusive
\tw rate is computed, including 
${\cal O}(1 / \log  m_t^2 / m_b^2 )$ corrections, and when
combined with
detailed Monte Carlo simulations including the top and $W$ decay
products, indicate that the \tw single top process may be extracted from
the considerable \ttbar and \wwj backgrounds at low luminosity runs
of the LHC.
\\[0.4cm]
\end{abstract}
\end{center}

\newpage
\setcounter{footnote}{0}
\renewcommand{\thefootnote}{\arabic{footnote}}

\section{Introduction}
\indent \indent
\label{intro}

The discovery of the top quark at the Fermilab Tevatron 
\cite{topdisc} completes
the fermionic sector of the Standard Model (SM) of particle physics.
However, many questions regarding the top remain, and require further
investigation to be settled.  The primary question is whether or not the
top is ``just another quark'', or if its large mass is
indicative that it is something more.  The top's enormous mass may be
a clue that it plays a special role in the Electroweak symmetry
breaking (EWSB), and many of the proposed extensions of the SM
explain the large top mass by allowing the top 
to participate in modified or nonstandard dynamics
\cite{extensions}, connected to the physics which provides the
mass of the $W$ and $Z$ bosons.
This intriguing hypothesis is best
explored by careful study of top observables.  In particular, single top
production, as a measure of the top quark's electroweak interactions,
is an excellent place to look for new physics related to the EWSB
\cite{mythesis}.  

Single top production proceeds through three distinct
modes at a hadron collider.  The choice of the word ``mode'',
as opposed to ``sub-process'' is motivated by the fact that
each process has different initial and final states, and thus
they are in principle separably measurable.
The $t$-channel $W$-gluon fusion
mode \cite{tchannel,willendicus,dthesis,newt}
involves the exchange of a space-like $W$ boson between a
light quark, and a bottom ($b$) quark inside the incident hadrons,
resulting in a jet and a single top quark.  Its rate is rather
large at both the Tevatron
and the LHC.  The $s$-channel
$W^*$ mode \cite{schannel}
involves production of an off-shell, time-like $W$ boson,
which then decays into a top and a bottom quark.  It has a relatively
large rate at the Tevatron, but is comparatively small at the LHC because
it is driven by initial state anti-quark parton densities.
Finally, the \tw mode of single top production involves an initial
state $b$ quark emitting a (close to) on-shell $W^-$ boson,
resulting in a \tw final state.  Because of the massive particles
in the final state, this mode has an extremely small rate at the Tevatron,
but is considerable at the LHC where more partonic energy is available.

Each mode has rather distinct event kinematics, and thus are
potentially observable separately from each other 
\cite{dthesis}.
In fact it has been shown \cite{mythesis,new} that each mode is
sensitive to different types of new physics, with the \tw mode
distinct in that it is sensitive only to physics which directly
modifies the $W$-$t$-$b$ interaction from its SM structure.
This distinction is a result of the fact that in this mode
both the top and the $W$ are directly observable, whereas
in the other two modes the $W$ bosons are virtual, and thus
those processes may receive contributions from exotic types of
charged bosons or FCNC operators involving the top.  Even without
invoking additional particles or FCNC interactions, 
the three modes provide complimentary information about the 
$W$-$t$-$b$ interaction by probing it in different regions of
momentum transfer.  The $t$-channel and $s$-channel processes
study the vertex when the $W$ boson is space-like or time-like,
respectively, and the fermions are (approximately)
on-shell.  The \tw mode involves the interaction when
the $W$ boson is on-shell, and one of the fermion lines is
off-shell.  These considerations are compelling reasons
to examine the three modes of single top separately, to extract
the maximum possible information from single top production.

This article contains a detailed study of the \tw mode of
single top production with the intent to observe this process
for its own sake.  Previous analyses focused
on this process as a background to heavy Higgs decays ($h \ra W^+ W^-$)
\cite{twold}
or combined all three single top modes together into one signal
\cite{rus1,rus2}.  For the reasons stated above, it is also very 
important to study the
\tw mode independently from the other single top processes.
The presentation
is arranged as follows.  In Section~\ref{inclusive}, the
inclusive \tw rate is computed, including large logarithmic corrections
of ${\cal O}( 1/ \log  m_t^2 / m_b^2 )$.  In Section~\ref{mcstudy}
a Monte Carlo simulation of the signal and the major backgrounds is
discussed for the LHC collider, and it is demonstrated that the \tw
signal may be observed with about 1 \ifb of collected data.
A summary of the results are given in Section~\ref{conclusions}.
Finally, the helicity amplitudes for the
signal process at leading order (LO), 
including all decays are presented in the appendix.

\section{The inclusive \tw Rate}
\label{inclusive}

\subsection{Leading Order $2 \ra 2$ Contributions}
\indent \indent

\begin{figure}[t]
\epsfysize=1.5in
\centerline{\epsfbox{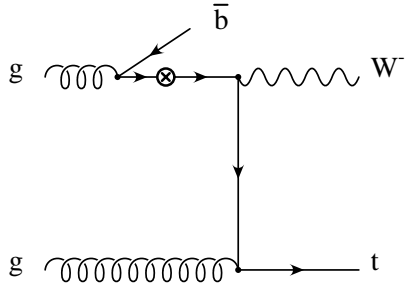}}
\caption{The underlying picture for production of \tw
at a $p p$ or $p \bar{p}$ collider.  The crossed
circle on the bottom quark line
anticipates the improvement of the perturbation
series by resumming the $g \ra b \, \bar{b}$ portion of the
interaction into a $b$ parton density.}
\label{twpicfig}
\end{figure}

Before turning to the details of how the \tw process may be observed
against the background, it is necessary to discuss the inclusive rate
of this process at a hadron collider.  As is the case with the
$t$-channel single top mode
\cite{willendicus}, the \tw rate involves finding a $b$
quark inside one of the incident (anti-) protons.  The underlying
picture is actually that a gluon ($g$) in one of the protons splits into
a $b \, \bar{b}$ pair, with one of these bottom quarks taking part in
the hard scattering, as is shown in Figure~\ref{twpicfig}.
In order to facilitate the discussion below, only \tw production will
be considered, and not $\bar{t} \, W^+~$ production.  It should
be clear how the remarks on single {\it top} production may also
be applied to single {\it anti-top} production with an appropriate
switch of 
$b \lra \bar{b}$, $t \lra \bar{t}$, and $W^\pm \lra W^\mp$.

One could imagine computing the 
inclusive \tw rate from a gluon-gluon
initial \mbox{state}, such as the one shown in Figure~\ref{twpicfig}.
However, this picture results in a perturbative expansion that is
spoiled by the kinematic region in which the produced $\bar{b}$
quark is approximately collinear with the incoming gluon, which
produces a contribution containing large logarithms 
of the form $\alphas \log m_t^2 / m_b^2$, which for $m_t \sim 175$
GeV, $m_b \sim 5$ GeV, and $\alphas \sim 0.1$ is over-all of order
1.  In fact, the $n$th order correction to the process always
contains a collinear piece which behaves as
$(\alphas \log m_t^2 / m_b^2)^n$, spoiling the perturbative description.

A convergent perturbative expansion is restored by resumming these
logarithms into a bottom quark parton distribution function (PDF)
\cite{bpdf}, which is unlike the usual light parton PDF's in that
it is perturbatively derived from the gluon distribution function.  
Thus, the LO contribution to inclusive \tw production 
is best considered to arise from Feynman diagrams such
as those shown in 
Figure~\ref{twfig}\footnote{The helicity amplitudes for this
$2 \ra 2$ contribution, including the top and $W$ decays, can be
found in Appendix~\ref{app}}, which treat the bottom quark as
a parton.

\begin{figure}[t]
\epsfysize=1.5in
\centerline{\epsfbox{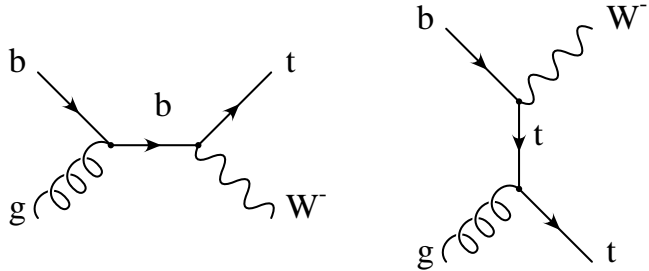}}
\caption{Feynman diagrams for the $t \, W^-$ mode of single top
production after resumming large logarithms into
a gluon PDF: $g \, b \ra t \, W^-$.}
\label{twfig}
\end{figure}

This reordering of the perturbation series improves its convergence,
and allows accurate estimation of the inclusive cross section.
In fact, this two particle to two particle ($2 \ra 2$) description
also represents the most important part of the \tw kinematics,
because the dominant kinematic configuration is one in which
the $\bar{b}$ is collinear with its parent gluon \cite{willendicus}.
However, since it effectively integrates out the $\bar{b}$
momentum, and approximates it as collinear with the parent
hadron, this $2 \ra 2$ process does not accurately describe the
situation when the $\bar{b}$ has large transverse momentum
($p_T$).  In this region, a description based on the 
original $2 \ra 3$ process is more appropriate, and since this
is the situation in which the $\bar{b}$ is not collinear with the
incoming gluon, it is well defined in perturbation theory.
Thus, we will see that while it is not possible to compute
kinematic distributions in the region in which the $p_T$ of the
$\bar{b}$ is small using standard perturbation theory, 
it is possible to compute the inclusive rate for the process
with $p_T^{\bar{b}} > p_T^{cut}$, so long as $p_T^{cut}$ is
chosen large enough that the perturbative description remains
valid.

\subsection{${\cal O}( 1 / \log  m_t^2 / m_b^2 )$ $2 \ra 3$ Corrections}
\indent \indent

Though the complete next-to-leading order
(NLO) QCD corrections are still underway,
the prediction for the inclusive \tw rate
may be improved over the LO result by including the
${\cal{O}}(1/ \log m_t^2 / m_b^2 )$ corrections coming from
Feynman diagrams such as those in Figure~\ref{twlogfig}.
These diagrams may be separated into three classes.
Figure~\ref{twlogfig}a contains a representative 
diagram which contains the
collinear ${\cal{O}}(1/ \log m_t^2 / m_b^2 )$ 
behavior under study.
Figure~\ref{twlogfig}b corresponds to a class of corrections that
include contributions which look like \ttbar production
followed by the decay $\bar{t} \ra W^- \, \bar{b}$.
The diagram of Figure~\ref{twlogfig}c contains a separate
${\cal O}(\alphas)$ correction that involves neither potentially
on-shell $\bar{t}$ quarks nor large logarithms.

There are two subtle points that must
be carefully dealt with when including these corrections.  The first
is that when the $b$ PDF was defined, the collinear contributions
from these diagrams were already resummed into what we called the
LO contribution.  Thus, in order to avoid double-counting this
collinear region one must subtract out this portion.
This may be expressed by writing the full cross section 
for $A \, B \ra t \, W^-$ as,
\bea
   \sigma_{tW} = \sigma^0(A \, B \ra t \, W^-)
   + \sigma^1(A \, B \ra t \, W^- \, \bar{b}) 
   - \sigma^{S}(A \, B \ra t \, W^- \, \bar{b}),
\eea
with the individual terms given by,
\bea
   \sigma^0(A \, B \ra t \, W^-) &=&
     \int  dx_1 \, dx_2 \left\{
     {f_{g / A}}(x_1, \mu) \, f_{b / B}(x_2, \mu) \,
     {\sigma}(b \, g \ra t \, W^-) \right. \\
   & & \left. + \; {f_{b / A}}(x_1, \mu) \, f_{g / B}(x_2, \mu)
     \, {\sigma}(g \, b \ra t \, W^-) \right\} \nonumber \\[0.2cm]
   \sigma^1(A \, B \ra t \, W^- \, \bar{b}) &=&
     \int dx_1 \, dx_2 \;
     {f_{g / A}}(x_1, \mu) \, f_{g / B}(x_2, \mu) \,
     {\sigma}(g \, g \ra t \, W^- \, \bar{b}) \nonumber \\[0.2cm]
   \sigma^{S}(A \, B \ra t \, W^- \, \bar{b}) &=& 
     \int dx_1 dx_2 \left\{
     {\tilde{f}_{b / A}}(x_1, \mu) \, f_{g / B}(x_2, \mu) \,
     {\sigma}(b \, g \ra t \, W^-) \right. \nonumber \\
   & & \left. + \; {f_{g / A}}(x_1, \mu) \, {\tilde{f}_{b / B}}(x_2, \mu)
     \, {\sigma}(g \, b \ra t \, W^-) \right\} , \nonumber
\eea
where $f_{i / H}(x, \mu)$ represents the parton distribution function
for parton $i$ carrying momentum fraction $x$ at scale $\mu$
to be found in hadron $H$.
The
``modified $b$ PDF'', ${\tilde{f}_{b / H}}$, 
contains the collinear logarithm\footnote{This result is derived
by including a small (but non-zero) bottom mass in 
the $2 \ra 3$ matrix elements in order to
regulate the collinear divergence occuring in the class of
diagrams represented by Figure~\ref{twlogfig}a.}
and splitting
function $P_{b \leftarrow g}$ convoluted with the gluon PDF,
\bea
   {\tilde{f}_{b / H}}(x, \mu) = \frac{ \alpha_S(\mu)}{2 \, \pi}
   \log \left( \frac{ \mu^2 }{ m_b^2 } \right) \int^1_x
   \frac{dz}{z} \left[ \frac{ z^2 + (1 - z)^2 }{2} \right]
   f_{g/H}\left(\frac{x}{z}, \mu \right).
\eea
By including this term, the collinear behavior in 
$\sigma^1 (A B \ra t \, W^-)$, which was already implicitly
included in $\sigma^0 (A B \ra t \, W^-)$, is removed.
Thus, the problem of double-counting
the collinear region is resolved.

\begin{figure}[t]
\epsfysize=1.7in
\centerline{\epsfbox{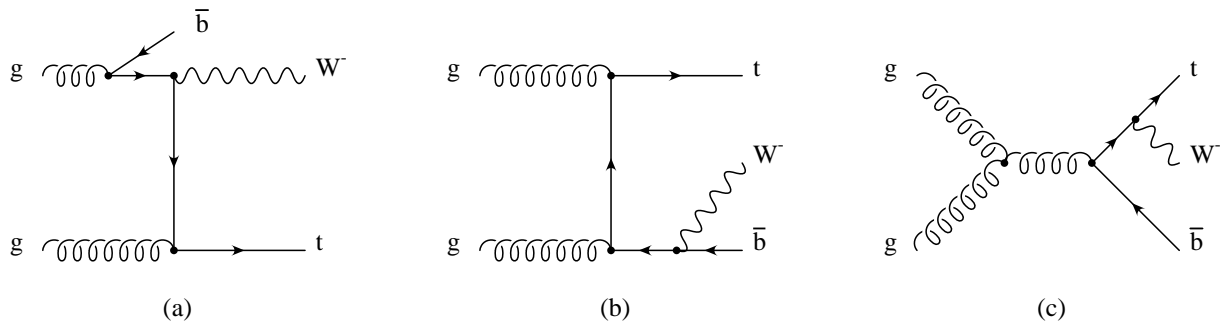}}
\caption{Representative Feynman diagrams for 
corrections to the $t \, W^-$ mode of single top
production corresponding to (a) large $\log$ corrections associated with
the $b$ PDF, (b) LO $t \, \bar{t}$ production followed by the LO
decay $\bar{t} \ra W^- \, \bar{b}$, and (c) pure $\alphas$ corrections.}
\label{twlogfig}
\end{figure}

The second subtle point
in evaluating the large $\log$ contributions is that they contain
contributions such as those found in Figure~\ref{twlogfig}b
that correspond to LO $g \, g \ra t \, \bar{t}$ 
production followed by the LO
decay $\bar{t} \ra W^- \, \bar{b}$.  This expresses the fact that as
one considers higher orders in perturbation theory, the distinction
between $t \, \bar{t}$ production and various types of single top
production is blurred.  However, when considering quantities that are
properly defined, these corrections are small, and there is no problem
distinguishing these processes.  As a matter of book keeping,
the corrections to $t \, W^-$ production involving an on-shell $\bar{t}$
are more intuitively
considered a part of the LO $t \, \bar{t}$ rate, and thus it
is important to subtract them out to avoid double counting in this
kinematic region. This may be done by noting that in the region
where the invariant mass of the $\bar{b} \, W^-$ system, $M_{Wb}$,
is close to the top mass, the behavior of the partonic cross
section ${\sigma}(g g \ra t \, W^- \bar{b})$ may be expressed,
\bea
   \frac{d{\sigma}}{dM_{Wb}}(g \, g \ra t \, W^- \, \bar{b}) &=&
   \sigma^{LO}(g \, g \ra t \, \bar{t}) \, 
   \frac{ m_t \, \Gamma^{LO}(\bar{t} \ra W^- \bar{b}) }
   {\pi \, [(M_{Wb}^2 - m_t^2)^2 + m_t^2 \, \Gamma_t^2]} \\
   &=& \sigma^{LO}(g \, g \ra t \, \bar{t}) \,
   \frac{ m_t \, \Gamma_t \, BR(\bar{t} \ra W^- \bar{b}) }
   {\pi \, [(M_{Wb}^2 - m_t^2)^2 + m_t^2 \, \Gamma_t^2]} \nonumber \\[0.3cm]
   &\ra& \sigma^{LO}(g \, g \ra t \, \bar{t}) \,
   BR(\bar{t} \ra W^- \bar{b}) \, \delta (M_{Wb}^2 - m_t^2) \nonumber
\eea
where $\sigma^{LO}(g \, g \ra t \, \bar{t})$ and 
$\Gamma^{LO}(\bar{t} \ra W^- \bar{b})$ are the LO cross section and
partial width, $\Gamma_t$ is the inclusive top decay width, and
$BR$ denotes the branching ratio.
The last distribution identity holds in the limit $\Gamma_t \ll m_t$.
Having identified this LO on-shell piece, it may now be simply subtracted
from ${\sigma}(g \, g \ra t \, W^- \, \bar{b})$.
The advantage to this formulation of the subtraction is that
by taking the narrow width limit, one removes all of the on-shell 
$\bar{t}$ contribution.
The interference terms between one of the on-shell $\bar{t}$ amplitudes 
and an amplitude without an on-shell $\bar{t}$ involve a Breit-Wigner
propagator of the form, $( M_{Wb}^2 - m_t^2 + i \, m_t \Gamma_t )^{-1}$,
which in the limit of small $\Gamma_t$, may be expressed as a 
principle valued integral in $M_{Wb}$.
Following this prescription,
and choosing a canonical scale choice
of $\mu_0 = \sqrt{s}$, where $s$ is the invariant mass of the
incoming partons,
leads to a large $\log$ correction to the
$t \, W^-$ rate of $-9.5\%$ at the LHC, which is consistent with
previous experience from the $W$-gluon fusion mode 
of single top production \cite{dthesis}.

This problem of the on-shell top was dealt with
another way in \cite{rus2}, where
a cut was applied on $M_{Wb}$, to exclude the region of
$|M_{Wb} - m_t| \leq 3 \, \Gamma_t$.
Following this prescription, one finds a much larger correction
of about $+50\%$
to the $t \, W^-$ rate at the LHC.  
However, this is misleading because the large
corrections are mostly coming from the region where the $\bar{t}$ is
close to on-shell (though still at least 3 top widths away).  In other
words, the large positive correction comes from the tails of the Breit-Wigner
distribution for on-shell $\bar{t}$ production.  This can be simply understood
by taking the prescription in \cite{rus2}
and varying the cut by increasing
the interval about the on-shell $\bar{t}$ region that is excluded.  
One finds that the correction computed in this way varies quite strongly with
the cut, and reproduces the subtraction method we have employed for
the cut $|M_{Wb} - m_t| \leq 12 \, \Gamma_t$.  
A theoretical
advantage of the subtraction method employed here
is that when one determines the
$t \, \bar{t}$ and $t \, W^-$ rates, one would like to actually fit the
data to the sum of the two rates, and thus the subtraction method
allows one to simply separate this sum into the two contributions
without introducing an arbitrary cut-off into the definition
of the separation.

Even if one were to use a cut-off to effect the separation, there is
a further problem in employing
the cut $|M_{Wb} - m_t| \leq 3 \, \Gamma_t$
to remove on-shell $t \, \bar{t}$ production.
This is that from a purely
practical point of view $3 \, \Gamma_t \sim 4.5 \,$ GeV, which is much smaller
than the expected jet resolution at the Tevatron or LHC.  Thus, it is not
experimentally possible to impose this definition of the separation between
$t \, W^-$ and $t \, \bar{t}$.  A more realistic resolution is
about 15 GeV \cite{lhcjetres}, which corresponds to a subtraction of
$|M_{Wb} - m_t| \leq 10 \, \Gamma_t$.  As we have seen above, this 
choice of the $M_{Wb}$ cut agrees rather well with
our subtraction method result.

In Tables~\ref{twtab} and \ref{twtab2} can be found
the LO rate (including the large $\log$ corrections
described above) of $t \, W^-$ production at the Tevatron
and LHC, for various choices of $m_t$,
the CTEQ4L \cite{cteq4} and MRRS(R1) \cite{mrrs} PDF's, 
and 3 choices of factorization scale,
with the canonical scale choice set to $\mu_0 = \sqrt{s}$.
These results assume\footnote{The inclusion of non-zero
$V_{ts}$ and $V_{td}$ have a negligible effect on the
cross section, because these parameters are required by
low energy processes to be extremely small \cite{pdg}.}
$V_{tb} = 1$, $V_{ts}, V_{td} = 0$ and no
decay branching ratios are included.
At both Tevatron and LHC, the rate for $\bar{t} \, W^+$ production
is equal to the rate of $t \, W^-$ production, and thus the
sum of the two rates may be obtained by multiplying the cross
sections by two.
From these results, we see that
varying the scale by a factor of two produces a variation in the
resulting cross section of about $\pm  25 \%$ at the Tevatron
and $\pm 15 \%$ at the LHC.
This large scale dependence signals
the utility in having a full NLO (in $\alpha_S$)
computation of this process in order to have
a more theoretically reliable estimate for the cross section.

It is further interesting to examine the dependence of the result
on the choice of PDF, as the \tw process is sensitive to the
gluon density, which is parameterized differently by the CTEQ4 and
MRRS PDF's.  This
provides one with an estimate of the
the uncertainty due to the difference in PDF parameterization.  
Unlike the light quark distributions,
which are rather well determined by deeply inelastic scattering (DIS)
data, the gluon PDF is relatively poorly known at large momentum
fraction, and thus the two parameterizations can lead to rather
different results for the \tw cross section.
For $m_t = 175$ GeV, comparing at the canonical scale choice,
a variation of $\pm 6\%$ at the Tevatron and $\pm 8\%$ at the LHC
results when moving from either the CTEQ4L or MRRS(R1) result
to the mean of the two.  However, care must be taken in drawing
firm conclusions from this because these two sets of PDF's are
extracted from a very similar collection of experimental data.
Thus, this estimate does not accurately reflect 
the uncertainty in the PDF's coming from the uncertainties 
in the data from which they are derived.

\begin{table}[t] 
\begin{center} 
\begin{tabular}{lccccccc}
 & \multicolumn{3}{c}{CTEQ4L} & 
\multicolumn{3}{c}{MRRS(R1)} & 
 \\
$m_{t}$ (GeV) & $\mu = \mu_0 / 2$ & $\mu = \mu_0$ &
$\mu = 2 \mu_0$ & $\mu = \mu_0 / 2$ & 
$\mu = \mu_0$ & $\mu = 2 \mu_0$ & $\sigma_{tW}^{(mean)}$
\\[0.2cm] \hline \hline \\
170 & 0.0645 & 0.0505 & 0.0405 & 0.0760 & 0.0580 & 0.0460 & 0.0545 \\
171 & 0.0630 & 0.0490 & 0.0395 & 0.0740 & 0.0565 & 0.0445 & 0.0530 \\
172 & 0.0610 & 0.0480 & 0.0385 & 0.0720 & 0.0550 & 0.0435 & 0.0515 \\
173 & 0.0595 & 0.0465 & 0.0375 & 0.0700 & 0.0530 & 0.0420 & 0.0500 \\
174 & 0.0575 & 0.0450 & 0.0365 & 0.0680 & 0.0515 & 0.0410 & 0.0485 \\
175 & 0.0560 & 0.0440 & 0.0355 & 0.0660 & 0.0500 & 0.0395 & 0.0470 \\
176 & 0.0545 & 0.0425 & 0.0345 & 0.0640 & 0.0490 & 0.0385 & 0.0460 \\
177 & 0.0530 & 0.0415 & 0.0335 & 0.0620 & 0.0475 & 0.0375 & 0.0445 \\
178 & 0.0515 & 0.0405 & 0.0325 & 0.0600 & 0.0460 & 0.0365 & 0.0435 \\
179 & 0.0500 & 0.0390 & 0.0315 & 0.0585 & 0.0445 & 0.0355 & 0.0420 \\
180 & 0.0485 & 0.0380 & 0.0305 & 0.0570 & 0.0435 & 0.0345 & 0.0410 \\
181 & 0.0475 & 0.0370 & 0.0300 & 0.0555 & 0.0425 & 0.0335 & 0.0400 \\
182 & 0.046  & 0.0360 & 0.029  & 0.0540 & 0.0410 & 0.0325 & 0.0385 \\
\hline \hline 
\end{tabular}
\end{center}
\caption{The LO (with ${\cal O}( 1 / \log m_t^2 / m_b^2)$
corrections)
rates of $b \, g \ra t \, W^-$ (in pb) at the Tevatron Run~II.
The rate of $\bar{t}$ production is equal to the rate of $t$
production.}
\label{twtab}
\vspace*{1.0cm}
\end{table}

\begin{table}[t]
\begin{center} 
\begin{tabular}{lccccccc}
 & \multicolumn{3}{c}{CTEQ4L} & 
\multicolumn{3}{c}{MRRS(R1)} & 
 \\
$m_{t}$ (GeV) & $\mu = \mu_0 / 2$ & $\mu = \mu_0$ &
$\mu = 2 \mu_0$ & $\mu = \mu_0 / 2$ & 
$\mu = \mu_0$ & $\mu = 2 \mu_0$ & $\sigma_{tW}^{(mean)}$
\\[0.2cm] \hline \hline \\
170 & 33.0 & 28.2 & 24.5 & 39.0 & 33.0 & 28.4 & 30.6 \\
171 & 32.2 & 27.5 & 24.0 & 38.3 & 32.5 & 27.9 & 30.0 \\
172 & 31.6 & 27.1 & 23.6 & 37.6 & 31.8 & 27.4 & 29.4 \\
173 & 31.1 & 26.6 & 23.1 & 38.0 & 31.3 & 26.9 & 28.9 \\
174 & 30.5 & 26.1 & 22.7 & 36.2 & 30.7 & 26.4 & 28.4 \\
175 & 29.9 & 25.6 & 22.2 & 35.4 & 30.1 & 26.0 & 27.9 \\
176 & 29.4 & 25.2 & 21.8 & 34.8 & 29.6 & 25.5 & 27.4 \\
177 & 28.9 & 24.7 & 21.5 & 34.2 & 28.9 & 25.0 & 26.8 \\
178 & 23.3 & 24.2 & 21.1 & 33.6 & 28.4 & 24.6 & 26.3 \\
179 & 27.8 & 23.7 & 20.7 & 33.0 & 27.9 & 24.1 & 25.8 \\
180 & 27.2 & 23.3 & 20.3 & 32.4 & 27.4 & 23.7 & 25.4 \\
181 & 26.8 & 22.9 & 20.0 & 31.8 & 26.9 & 23.2 & 24.9 \\
182 & 26.3 & 22.5 & 19.6 & 31.2 & 26.4 & 22.9 & 24.5 \\
\hline \hline 
\end{tabular}
\end{center}
\caption{The LO (with ${\cal O}( 1 / \log (m_t^2 / m_b^2)$
corrections) rates for
$b \, g \ra t \, W^-$ (in pb) at the LHC.
The rate of $\bar{t}$ production is equal to the rate of $t$
production.}
\label{twtab2}
\end{table}

\clearpage
\section{Extracting \tw from the Background}
\indent \indent
\label{mcstudy}

In this section we present the results of a Monte Carlo study illustrating
how one may extract the \tw signal from the relatively large
backgrounds.  A top mass of $m_t = 175$ GeV
is assumed.
The \tw signature consists of the decay products of an on-shell top 
quark and $W^-$ boson.  The SM top decay is $t \ra W^+ \, b$, and
the $W$ bosons may decay either into quarks or leptons.  While the
hadronic $W$ decays are dominant (with branching ratio 6 / 9), the
leptonic decays are generally cleaner, and energetic leptons provide
an excellent trigger at a hadron collider.  Thus, it is necessary to
consider at least one of the $W$ bosons decaying into leptons.
This study will be confined to the case where both $W$'s decay
into leptons, $W \ra \ell \nu$ with $\ell = e, \mu$.
This avoids potentially huge QCD backgrounds
involving production of one $W$ boson and two or more jets
whose invariant mass happens to lie close to $m_W$.
Thus, the signature contains two relatively hard charged leptons,
missing transverse energy ($\not p_T$) from the unobserved
neutrinos, and a bottom quark from the single top decay.

\begin{figure}[t]
\epsfysize=1.5in
\centerline{\epsfbox{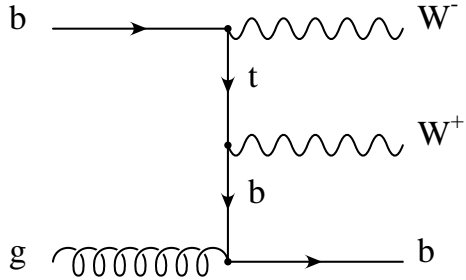}}
\caption{Illustrative Feynman diagram for the irreducible background,
\wwb production, involving an off-shell top quark.  The leptonic
decay modes of the $W$ bosons are not indicated here.}
\label{wwbfig}
\end{figure}

The primary background comes from processes which
contain two weak bosons and a bottom quark.  This includes
partonic processes such as 
continuum \wwb (shown in Figure~\ref{wwbfig})
which involve off-shell top exchange.  In fact, this process
is interesting in its own right as a further test of the $W$-$t$-$b$
interaction.  However, the lack of an on-shell top quark will
make this process quite difficult to distinguish from the backgrounds.
There are also important fake backgrounds from processes that are similar
to the \tw process, and may be misidentified as such.  This includes
\ttbar production, when one of the bottom quarks from a top decay
is unobserved because it falls outside of the detector coverage.
As the inclusive \ttbar rate is more than an order of magnitude
larger than the inclusive \tw rate, this is potentially a serious
problem in discovering the \tw process.  Another class of potentially
important backgrounds is \wwj production (where $j$ is a light quark
or a gluon) which are problematic because of the small but non-zero
probability that $j$ may be misidentified as jet containing a bottom
quark.  It should be noted that this set of backgrounds also includes
$Z \, Z \, j$ production, with one $Z$ decaying into charged
leptons, and the other into neutrinos.
As we shall see, the \wwj and \wwb backgrounds are much smaller
than either the \tw signal or the \ttbar background.  This is easily
understood from the fact that the \wwb and \wwj backgrounds are
$2 \ra 3$ production processes of ${\cal O}(\alphas \, \alpha_W^2)$
(followed by $W$ decays),
whereas the signal and \ttbar background are $2 \ra 2$ production
processes of ${\cal O}(\alphas \, \alpha_W)$ and ${\cal O}(\alphas^2)$
respectively, followed by top decays with unit branching ratio 
and the same $W$ decays.

The \tw signal, \wwb, and \wwj backgrounds are
simulated at the parton level, at LO in the strong and weak
couplings.  
The matrix elements for the \tw process, including
all of the decays have been computed at the helicity amplitude
level, and are presented in Appendix~\ref{app}.  Matrix elements
for the \wwb and \wwj processes have been obtained from the
MADGRAPH code \cite{madgraph}.
The \ttbar background is simulated at LO, using the ONETOP code
\cite{onetop}.  For all processes, a scale choice of the partonic
energy, $\mu = \sqrt{s}$ is chosen, and the CTEQ4L PDF is used
with LO running of the strong coupling\footnote{The $\Lambda_{QCD}$
with 5 active flavors appropriate for the CTEQ4L fit is 181 MeV.}.
Though NLO results for \ttbar production are currently available
\cite{nlott}, we use the LO rates in order to make a fair
comparison between this background, the signal, and the other
backgrounds, for which NLO results are currently unknown.

In order to be considered observable, the charged leptons and
$b$ jet are required to have $p_T \geq 15$ GeV, and to lie in
the central region of the detector, 
with rapidity $|\eta| \leq 2$.  In order to
be resolvable as distinct objects, the charged lepton and jet
are required to have a cone separation $\Delta R \geq 0.4$,
where $\Delta R = \sqrt{ \Delta \eta^2 + \Delta \phi^2}$,
with $\Delta \phi$ the separation in the azimuthal angle,
and $\Delta \eta$ the difference in pseudorapidity.  In the
second column of Table~\ref{eventstab} can be found
the number of events
passing these cuts for signal and backgrounds at the LHC,
assuming 20 \ifb of integrated luminosity.

\begin{table}[t] 
\label{eventstab}
\begin{center} 
\begin{tabular}{lccc}
Process~~~~ & ~~~Acceptance Cuts~~~ & ~~~$p_T^{\bar{b}} \leq 15$ GeV~~~ 
            & ~~~$b$-tagging~~~ \\
\hline \hline
\tw     & 25250  & 13535 & 8121 \\
\wwb    & 114    & 114   & 68   \\
\wwj    & 2460   & 2460  & 25   \\
\ttbar  & 270495 & 15074 & 9044 \\
\hline \hline
\end{tabular}
\end{center}
\caption{The number of signal \tw events as well as the major
backgrounds, for 20 \ifb of integrated luminosity at the LHC
after applying cuts described in the text.}
\vspace{0.3in}
\end{table}

In order to suppress the large \ttbar background, it is
desirable to exclude events with more than one $b$ quark
at high transverse momentum.  The estimation of the 
effect of this cut on the \tw signal is somewhat delicate,
because, as was discussed in Section~\ref{inclusive},
the kinematics of the $\bar{b}$ in the signal events are
not calculable in usual formulation of perturbative QCD.
However, the effect of the cut may nonetheless be computed
using a technique developed for the $W$-gluon
fusion process \cite{newt}.  The
inclusive \tw rate computed in Section~\ref{inclusive}
represents the \tw rate summed over all $p_T$ of the $\bar{b}$.
Further, one can use the $2 \ra 3$ description with a
sufficiently strong cut on the $p_T$ of the $\bar{b}$ to
reliably obtain the cross section when the $\bar{b}$ has
a large transverse momentum.  Since the inclusive rate is
equal to the sum of the rate below the $p_T$ cut and the
rate above the $p_T$ cut, one thus computes the signal
cross section below the $p_T$ cut from,
\bea
   \sigma^{t \, W^-}( p_T^{\bar{b}} \leq p_T^{cut}) &=&
   \sigma^{t \, W^-}_{inclusive} - 
   \sigma^{t \, W^-}( p_T^{\bar{b}} \geq p_T^{cut}).
\eea
The effect of the cut $p_T^{\bar{b}} \leq 15$ GeV
on the signal\footnote{Since this cut
restricts the $p_T$ of the $\bar{b}$ to be small, we continue
to simulate the signal rate from the $2 \ra 2$ process
(followed by decays)
described in the Appendix, which makes the approximation
$p_T^{\bar{b}} = 0$.}
and backgrounds is shown in the third column of 
Table~\ref{eventstab}.  As is indicated, this cut
is extremely effective at suppressing the \ttbar background,
drastically  reducing it by about $94\%$, 
while eliminating only about $47\%$ of the signal rate.

The \wwj background has been simulated using the exact
matrix elements for $q \, \bar{q} \ra W^+ \, W^- \, g$ and
$q \, g \ra W^+ \, W^- \, q$,
which is appropriate for high energy jets.  However, it is
expected that this background may receive further contributions
from initial and final state showering, and fragmentation,
which have not been included in our estimate.  Thus, in order to
be conservative, we include $b$-tagging in our search to
select events containing a bottom quark.  The $b$-tagging
efficiency is estimated by assuming a $60\%$ probability of
correctly identifying a bottom quark passing the acceptance
cuts described above.  The probability of mis-tagging a
light quark\footnote{The probability of misidentifying a charm
quark as a bottom is closer to $15\%$.
However, as the \wwc portion of the already small \wwj background
is only about $7\%$, this is of negligible significance.}
or gluon is assumed to be $1\%$.  The effect
of this requirement is indicated in the final column of
Table~\ref{eventstab}.

One may now estimate the ability of the LHC to study the \tw
process.  The significance of the signal over the background
is defined as the number of signal events divided by the 
square-root of the number of background events.  As the total
number of events satisfying our search criterion is large,
this approximation based on Gaussian statistics is justified.
From the final column of Table~\ref{eventstab}, the significance
at the LHC with 20 \ifb of integrated luminosity 
is seen to be 84.9, indicating that even with a relatively 
small amount of data the LHC will be able to identify the
signal from the background.  An alternative presentation is
that the LHC will observe the \tw mode of single top production
at the $5\sigma$ level with less than 1 \ifb of data.
In comparison, at the Tevatron Run~II
with 2 \ifb, it is found that after
applying the acceptance and $p_T^{\bar{b}}$ cuts,
the signal rate is expected to be slightly less than 1 event, thus
indicating that another search strategy must be employed
in order to observe \tw at the Tevatron.

\newpage
\section{Conclusions}
\indent \indent
\label{conclusions}

This article is the first proposal to separately study the 
\tw mode of single top production for its own sake.  The inclusive
rate has been computed, including the large $\log$ corrections from
the definition of the bottom quark PDF.  This procedure
requires some attention because of potential problems with double-counting
both the region where the $\bar{b}$ quark is collinear with the
incident gluon, and with \ttbar production.  After appropriate subtractions,
the corrections to the \tw rate are determined, and it is found that the
net correction is about $-10\%$ at the LHC.

A full parton-level event simulation has been completed
at the LHC, including
the decay products from the top and $W$ bosons.  The decay mode
in which both $W$ bosons decay into leptons has been identified as
a signature with a relatively small QCD jet background, though
other decay modes are also potentially interesting.
The \ttbar rate
is found to be a serious background, but it may be reduced
by applying simple cuts to extract the signal from the background.
After applying these cuts, it is found that a $5\sigma$ observation
of the \tw signal is possible at very low luminosities at the
LHC.  Turning this
around, one finds a statistical error of about $1\%$ in the 
measured cross section at the LHC with 20 \ifb.
This error is much smaller than the theoretical uncertainty of
$\pm 15\%$ from the scale dependence, which
motivates further work to include higher orders of
perturbation theory in the \tw rate.

If the top quark has indeed been given a special
role in the generation of masses, it is crucial that
its interactions be carefully studied in order to learn
what properties the underlying theory at high energies must
possess.  The deviations of the top interactions from the
SM predictions may represent the best clues on the nature of
the EWSB.
The \tw rate of single top production represents an opportunity
to learn about the top quark's weak interactions from a different
perspective than is afforded by the $W$-gluon fusion and $W^*$
modes, both because the top {\it and} $W$ are observed, thus allowing
one to study the $W$-$t$-$b$ coupling independently from the 
possibility of FCNC and heavy particles, and also to probe the
$W$-$t$-$b$ interaction in a different region of momentum transfer
from the other two channels.
It thus represents an essential means to determine
the properties of the top quark.

\section{Acknowledgements}
The author is grateful for invaluable conversations
and encouragement from C.--P. Yuan, without which
this project would not have been possible,
and to B. Harris for reading the manuscript.
This work was completed at Argonne National Laboratory,
in the High Energy Physics division and was supported in
part by the U.S. Department of Energy, High Energy Physics
Division, under Contract W-31-109-Eng-38.

\appendix
\section{Helicity amplitudes for 
$b \, g \ra t \, W^- \ra b e^+ \nu_e \mu^- \bar{\nu_\mu}$}
\label{app}
\indent \indent

In this appendix, the helicity amplitudes for the \tw process
are presented, including the decay products of the $t$ and $W$.
The helicity amplitudes for the other modes of single top production,
as well as the \ttbar background, may be found in \cite{dthesis}.
Our conventions for implementing the helicity amplitudes may be
found in Appendix A of that work, and are embodied in the 
two-component ket notation for the spinors,
\bea
|p_i +> \; = \; \sqrt{2 \, E_i} \,
   \left( \begin{array}{c} \cos \theta /2 \\
                          e^{i \phi} \sin \theta /2 \end{array} \right), 
   &\,&
|p_i -> \; =  \; \sqrt{2 \, E_i} \,
   \left( \begin{array}{c} -e^{i \phi} \sin \theta /2 \\
                          \cos \theta /2 \end{array} \right) \\
\nonumber
\eea
where $E_i$, $\theta$, and $\phi$ refer to the energy, polar angle,
and azimuthal angles of four-momentum $p_i$.  A further notational
convenience is to introduce $2 \times 2$ Dirac matrices
$\gamma^{\pm}_\mu = (1, \pm \vec{\sigma})$, from which
follows the ``slash'' notation $\not p^\pm = p^\mu \gamma^\pm_\mu$.
Particle momenta are indicated by the hatted particle label, i.e.,
$p_{e^+} = \hat{e}^+$.

Purely for the purposes of labelling, we distinguish the decay
products of the $W^-$ boson from the $W^+$ boson by assuming the
decays,
\bea
   W^- \ra \mu^- \, \bar{\nu}_\mu, \\
   t \ra b \, W^+; \: W^+ \ra e^+ \, \nu_e . \nonumber
\eea
The other modes of the $W$ decays may be easily obtained from this
choice by including the appropriate sum over colors for each set of
$W$ decay products in the final (cross section) result.  Namely,
each hadronic $W$ decay into quarks multiplies the cross section by
$N_C = 3$ compared to the leptonic modes presented here.

The results do not include the (negligible) masses of any
fermions other than the top.  This assumption of massless fermions
has a profound effect on the helicity structure of the matrix elements
because of the left-chiral structure of the $W$ interaction.
The net result is that only the particular helicity structure
$\lambda_{b^i} = -1$, $\lambda_{\mu^-} = -1$, 
$\lambda_{\bar{\nu}_\mu} = +1$, $\lambda_{e^+} = +1$,
$\lambda_{\nu_e} = -1$, $\lambda_{b^f} = -1$ 
has a non-zero contribution,
where $\lambda_n = \pm 1$ when the spin polarization of particle
$n$ is along (against) its direction of motion.  
$b^i$ denotes the initial state $b$ quark, while
$b^f$ denotes the final $b$ resulting from the $t$ decay.
The only
remaining polarization to be specified is that of the initial
gluon.  
Denoting the initial gluon spin as $\pm$ for right (left)
handed gluons, the two amplitudes may be expressed as,
\bea
& & \\[0.4cm]
{\cal M}_1(+) &=& \left( \frac{\sqrt{2} \, g_W^4 \, g_S (2 E_g)^{-1}}
                       {(\hat{t}^2_1 - m_t^2)
                        ({\hat{W}_2}^2 - m_W^2 + i m_W \Gamma_W)
                        ({\hat{W}_1}^2 - m_W^2 + i m_W \Gamma_W)
                        ({\hat{t}_2}^2 - m_t^2 + i m_t \Gamma_t)} \right)
  \nonumber \\ & & \times
                    <\hat{b}^f-|\hat{\nu}_e+> \, 
                    <\hat{\bar{\nu}}_\mu+|\hat{b}^i-> 
  \nonumber \\
  & & \times \left( m_t^2 <\hat{e}^++|\hat{g}+> \, <\hat{g}-|\hat{\mu}^-+> 
                   - <\hat{e}^++|\not \hat{t}_2^{-}|\hat{g}+> \,
                     <\hat{g}-|\not \hat{t}_1^{-}|\hat{\mu}^-+> \right)
  \nonumber \\[0.8cm]
{\cal M}_1(-) &=& \left( \frac{-\sqrt{2} \, g_W^4 \, g_S (2 E_g)^{-1}}
                       {(\hat{t}^2_1 - m_t^2)
                        ({\hat{W}_2}^2 - m_W^2 + i m_W \Gamma_W)
                        ({\hat{W}_1}^2 - m_W^2 + i m_W \Gamma_W)
                        ({\hat{t}_2}^2 - m_t^2 + i m_t \Gamma_t)} \right)
  \nonumber \\ & & \times
                    <\hat{b}^f-|\hat{\nu}_e+> \, 
                    <\hat{\bar{\nu}}_\mu+|\hat{b}^i-> \nonumber \\
  & & \times \left( m_t^2 <\hat{e}^++|\hat{g}-> \, <\hat{g}+|\hat{\mu}^-+> 
                   - <\hat{e}^++|\not \hat{t}_2^{-}|\hat{g}-> \,
                     <\hat{g}+|\not \hat{t}_1^{-}|\hat{\mu}^-+> \right)
  \nonumber
\eea
\bea
{\cal M}_2(+) &=& \left( \frac{-\sqrt{2} \, g_W^4 \, g_S (2 E_g)^{-1} }
                         {\hat{b}^2_1
                        ({\hat{W}_2}^2 - m_W^2 + i m_W \Gamma_W)
                        ({\hat{W}_1}^2 - m_W^2 + i m_W \Gamma_W)
                        ({\hat{t}_2}^2 - m_t^2 + i m_t \Gamma_t)} \right)
  \nonumber \\ & & \times
                    <\hat{b}^f-|\hat{\nu}_e+> \, 
                    <\hat{e}^++|\not \hat{t}_2^{-}|\hat{\mu}^-+> \,
                    <\hat{\bar{\nu}}_\mu+|\not \hat{b}_1^{-}|\hat{g}+> \,
                    <\hat{g}-|\hat{b}^i->
  \nonumber \\[0.8cm]
{\cal M}_2(-) &=&   \left( \frac{\sqrt{2} \, g_W^4 \, g_S (2 E_g)^{-1} }
                         {\hat{b}^2_1
                        ({\hat{W}_2}^2 - m_W^2 + i m_W \Gamma_W)
                        ({\hat{W}_1}^2 - m_W^2 + i m_W \Gamma_W)
                        ({\hat{t}_2}^2 - m_t^2 + i m_t \Gamma_t)} \right)
  \nonumber \\ & & \times
                    <\hat{b}^f-|\hat{\nu}_e+> \, 
                    <\hat{e}^++|\not \hat{t}_2^{-}|\hat{\mu}^-+> \,
                    <\hat{\bar{\nu}}_\mu+|\not \hat{b}_1^{-}|\hat{g}-> \,
                    <\hat{g}+|\hat{b}^i-> \,
  \nonumber
\eea
where,
\bea
\hat{W}_1 &=& \hat{e}^+ + \hat{\nu}_e, \\
\hat{W}_2 &=& \hat{\mu}^- + \hat{\bar{\nu}}_\mu, \nonumber \\
\hat{t}_1 &=& \hat{b}^i - \hat{W}_2, \nonumber \\
\hat{t}_2 &=& \hat{b}^f + \hat{W}_1, \nonumber \\
\hat{b}_1 &=& \hat{b}^i + \hat{g},  \nonumber
\eea
and $E_g$ is the gluon energy, $g_S$ is the strong coupling constant,
and $g_W = e / \sin \theta_W$ is the weak coupling.  The matrix
elements squared, $|{\cal M}|^2$, should be summed over final
colors and averaged over initial colors, resulting in a factor of
$1/6$, and averaged over initial spins, giving a further factor of
$1/4$.  As was noted above, this color factor reflects the case
where both $W$ bosons decay into leptons.  A $W$ decay into quarks
further multiplies the cross section by $N_C =3$ for one hadronic
$W$ decay, and $N^2_C = 9$ if both $W$'s decay into hadrons.
The propagators for the (approximately) on-shell $W^\pm$ and top have
been expressed according to the Breit-Wigner prescription.


\end{document}